\documentclass[aps,chaos,preprint,preprintnumbers,amsmath,amssymb]{revtex4}

\usepackage{graphicx}
\usepackage{dcolumn}
\usepackage{bm}
\usepackage{epsfig}

\begin{document}

\title{Symbolic Synchronization and the Detection of Global Properties of 
Coupled Dynamics from Local Information}
\author{Sarika Jalan} \email{sjalan@mis.mpg.de}
\author {J\"urgen Jost} \email{jjost@mis.mpg.de}
\author{Fatihcan M. Atay} \email{atay@member.ams.org}
\affiliation{Max Planck Institute for Mathematics in the Sciences, 
 04103 Leipzig, Germany}

\date{\today}

\begin{abstract}We study coupled dynamics on networks using symbolic
dynamics. The symbolic dynamics is defined by dividing the state
space into a small number of regions (typically 2), and considering
the relative frequencies of the transitions between those
regions. It turns out that the global qualitative
properties of the coupled dynamics can be classified
 into three different phases based on the synchronization of 
the variables and the 
homogeneity of the symbolic dynamics. Of particular interest is
the {\it homogeneous unsynchronized 
phase} where the coupled dynamics is in a chaotic unsynchronized
state, but exhibits (almost)
identical symbolic dynamics at all the nodes in the network. We refer
to this dynamical behaviour as {\it symbolic synchronization}. 
In this phase, the local symbolic dynamics of any arbitrarily selected node
reflects global properties of the coupled dynamics, such as
qualitative behaviour of the largest
Lyapunov exponent and phase synchronization. 
This phase depends mainly on the  
network architecture, and only to a smaller extent on the local chaotic 
dynamical 
function. We present results for two model dynamics, iterations of the 
one-dimensional logistic
map and the two-dimensional  H\'enon map, as local 
dynamical function.
\end{abstract}  
\maketitle 
{\bf{Nonlinear dynamical elements
interacting with each other
can lead to synchronization or other types of coherent behaviour at
the system scale. 
Coupled map models are one of the most widely accepted models to understand 
these behaviours in systems from many diverse fields such as
physics, biology, ecology etc. Their important feature is that the
individual elements can already exhibit some complex behaviour, for
example chaotic dynamics. The question then is how to detect
coordination at larger scales beyond the simplest one,
synchronization. An important tool in the analysis of dynamical
systems are symbolic dynamics. 
We develop a new scheme of
symbolic dynamics that is based on the  
special partitions of the phase space which prevent the occurrence of
certain symbol sequences related to the characteristics of the dynamics.
In particular,
we report a new behaviour of coupled dynamics, which we refer to  as 
symbolic synchronization, i.e. synchronization of the nodes
at the coarse grained level,
whereas microscopically all elements behave differently. 
Through the framework of this symbolic dynamics, we detect various global 
properties of coupled dynamics on networks by using a scalar time series of any
randomly selected node. A decisive advantage of our method is that
the global properties are inferred by using a short time series, hence 
the method is computationally fast,  
does not depend on the size of the network, and is reasonably
robust against external
noise.}}

\section{Introduction}
In order to gain insights into the behaviour of real systems from many 
diverse fields  
ranging from chemical, physical and biological systems,
it is useful to identify model systems that on the one hand exhibit
essential dynamical features of those real world systems, but on the
other hand suppress individual details that are not really relevant
for the qualitative behaviour \cite{rev-CM}. 
Coupled map models have emerged as one
such paradigm \cite{rev-kaneko,SJ-REA-insa}. Here, we have a system of elements with 
identical local
dynamical functions. These elements are arranged in a network that
expresses their couplings so that the local dynamical iteration
depends not only on the own state of an element, but also on the ones
of its neighbour in the network. The inhomogeneities in the network
then translate into qualitative features of the global dynamics. While
such coupled map models already present an important simplification in
view of the complexities of real world dynamics, their behaviour can
nevertheless be sufficiently complicated and difficult to
analyze. Thus, it is important to identify parameters that allow for a
facile and robust detection of different qualitative states. One
needs a coarse grained description to analyze the complicated
time evolution of a chaotic dynamical system \cite{sym-book,sym-book2}. 
In doing so one inevitably 
simplifies the dynamics  a lot and some of the information are lost,
but the aim is that 
important invariants and robust properties of the dynamical systems
can be kept. Such a coarse graining means that we divide the possible
dynamical states of the system into finitely many discrete classes and
investigate the derived {\it symbolic dynamics} \cite{sym-book}. \\
Coarse graining of the time evolution of lower dimensional systems
have been studied at various levels \cite{sym-book2,sym-rev}, 
but symbolic dynamical studies of higher dimensional spatio-temporal chaotic
systems are rare and so far limited to coupled map lattices \cite{sym-CD}. 
In the present paper, continuing the approach developed in \cite{pap1}, 
we study symbolic dynamics of coupled map networks and demonstrate that they 
can serve the above purpose well. Thereby, we attempt to provide a general 
framework for coupled dynamics on networks.
                
In \cite{pap1}, we have studied
symbolic dynamics of a discrete dynamical iteration of a function
based on non-generating partitions.  We have shown two important uses of
this symbolic dynamics, namely,
distinguishing a chaotic iteration and a random iteration with the
same density distribution (this is related to the earlier
work on transition entropy \cite{kurths-TS} and \cite{PE1} on permutation 
entropy), and detecting 
synchronization in coupled dynamics on large networks.
In this paper we extend these studies and propose a
general method to investigate collective behaviour 
of coupled systems. 
Besides the applications mentioned in \cite{pap1}, we show further
applications of symbolic dynamics for detecting
various dynamical properties of
coupled dynamics in relation to structural parameters of the
underlying network.

We take coupled map network models as generic models to apply our
method. Chaotic coupled maps show rich spatio-temporal behaviour.
One  phenomenon that has received a lot of attention is 
synchronization, where different random or chaotic
units of a system behave
in unison
\cite{book-syn,phyreport-syn}. (For a selection of recent references, 
see also \cite{syn}.)
One application of our symbolic dynamics  is the detection of
synchronization in large complex systems.
Traditional methods for the detection of synchronization in  coupled
systems focus on the correlation analysis of the time series 
measured at  pairs of the nodes. 
In \cite{pap1} we have introduced a method based on symbolic dynamics, which
uses a short time series of 
any single arbitrarily selected node to detect global synchrony of all the 
units. 
In this paper we show that the same type of symbolic dynamics can be used 
as a measure of phase synchronization, a phenomenon
shown by coupled dynamics on  networks \cite{PS1}. 

In more detail, by our method we classify the coupled dynamics into
different states, depending
upon the synchronization of the nodes and the {\it homogeneity of the
symbolic dynamics} of the nodes. The most interesting phase is
the unsynchronized homogeneous phase, which refers to the state
where the local chaotic dynamics of the individual nodes are different, 
but the derived  symbolic dynamics of all 
the nodes are similar. We refer to this state as {\it symbolic
synchronization} of the nodes. Recently, the unsynchronized region of
coupled maps has been shown to have a fractal stationary density function 
\cite{jost-kiran}.
We show that, in this phase,
the transition probabilities 
of any randomly selected node reflect the qualitative information of the
largest Lyapunov exponent  ($\lambda_l$) and the phase synchronization 
of the coupled dynamics.
For the calculation of the  largest Lyapunov exponent we utilize only a 
short
time series, whereas traditional methods to calculate the largest Lyapunov
exponent
from a scalar time series require rather long time series and also
involve various computational complications \cite{lya-rev}. We point 
out, however, 
that -- as to be expected from such a simplistic reduction -- our symbolic 
dynamics gives only the qualitative behaviour of the
 Lyapunov exponent $\lambda_l$, but of course not its exact value.

The paper is organized as follows.
After an introductory section we introduce
the definitions of the different phases based on the symbolic 
dynamical properties 
in Section II. In Section III, we then present numerical
examples illustrating the behaviour of nodes in different phases. Mostly
we present results for homogeneous synchronized phase which is of main interest.
The key point is that the derived symbolic dynamics allows for the 
detection of the global properties
of coupled dynamics from local measurements, that is, we can infer global 
properties of the dynamical network by considering the symbolic dynamics at 
a single node. Section IV distinguishes different 
dynamical phases based on the network parameters. Section V describes the 
relation between symbolic dynamics 
and phase synchronization. Section VI discusses the coupled
H\'enon map. 

\section{Model and definition of symbolic dynamics}
We consider the dynamical system defined by the iteration rule
\begin{equation}
x(t+1)=f(x(t))
\label{iteration}
\end{equation}
where $t \in {\mathbb Z}$ is the discrete time and $f:S\rightarrow S$ is
a map on a subset $S$ of  ${\mathbb R}^{n}$.
Let $\{S_{i}:i=1,\ldots ,m\}$ be a partition
of $S$, i.e., a collection of mutually disjoint and nonempty subsets satisfying $\cup
_{i=1}^{m}S_{i}=S$. 
The symbolic dynamics corresponding to (\ref{iteration})
is the sequence of symbols $\{{\dots ,s_{t-1},s_{t},s_{t+1},\dots \}}$ where 
$s_{t}=i$ if $x(t)\in S_{i}$.
For the purposes of this paper, a useful partition is defined as follows.
Let $x=(x_{1},\dots,x_{n}) \in \mathbb{R}^n$, and
suppose the scalar $x_{n^\prime}$, $1\leq n^\prime\leq n$, is
available for measurement. For a given threshold value $x^{\ast} \in \mathbb{R}$,
define the sets
\begin{equation}%
\begin{array}
[c]{c}%
S_{1}=\{x\in S:x_{n^\prime}<x^{\ast}\}\\
S_{2}=\{x\in S:x_{n^\prime}\geq x^{\ast}\}
\end{array}
\label{part}%
\end{equation}
The value $x_{n^\prime}$ can be chosen to make the sets
$S_{1},S_{2}$ nonempty, in which case they form a non-trivial partition of $S$.
For this special partition, we use the two-symbol dynamics generated by
\begin{equation}
s_t= \left\{  \begin{array}{c}
                            \alpha \text{ if } x_n^{\prime}(t) < x^{\ast}  \\
                            \beta \text{ if } x_n^{\prime}(t) \ge x^{\ast}.
                 \end{array} \right .                                   
\label{sym}
\end{equation}
The symbolic dynamics depends only on the measurements $x_{n^{\prime}} $,
yielding a sequence of symbols determined by whether a measured value
exceeds the threshold $x^{\ast}$ or not.
Essentially any choice of the threshold $x^{\ast}$ will yield a
non-generating partition. For practical calculations using short time series,
however, a judicious choice of $x^{\ast}$ becomes important. We will address
this issue later in the paper (see section V).

We take the well known coupled map model \cite{CM},
\begin{equation}
x_i(t+1) = f(x_i(t)) + \frac{\varepsilon}{k_i} \sum_j w_{ij} g(x_j(t),x_i(t))
\label{coup-dyn}
\end{equation}
where $x_{i}(t)$ is the dynamical variable of the $i$-th node $(1 \le i \le N)$ 
at time $t$, $w$ is the adjacency matrix with elements $w_{ij}$
taking values between 0 and 1 depending upon the weight of the connection 
between $i$ and $j$, and $k_i$ is some normalization factor depending
on the node $i$, for example its degree. The function $f(x)$ defines the local 
nonlinear map and 
the function $g(x)$ defines the nature of the coupling between the nodes. 
In the first three sections, we present the results for the local
dynamics given by the logistic map $f(x) = \mu x (1-x)$ and 
coupling function $g(x_j(t),x_i(t))=f(x_j(t))-f(x_i(t))$. We take $\mu=4$,
for which logistic map exhibits chaotic behaviour with Lyapunov
exponent $\ln(2)$. The weight
$w_{ij}$ is simply one when nodes $i$ and $j$ are neighbours in the
undirected network, and 0 otherwise. In particular, the matrix $w$ is
symmetric; $k_i$ then is the degree of node $i$, as already
indicated. 

We evolve Equation~(\ref{coup-dyn}) starting from random initial conditions
and estimate the transition probabilities using time series of length 
$\tau=1000$. Note that the length of the time series is independent of the
size of the network. We calculate the transition probability $P(i, j)$
by the ratio $\sum_t n(s_t = i, s_{t+1} = j)/\sum_t n(s_t = i)$, where
$n$ is a count of the number of times of occurrence \cite{pap1}.

\section{Different states of the coupled dynamics}

We classify the coupled dynamics in three different categories 
based on the dynamical behaviour, and we show that how one category
differs from another based on some of the parameters of underlying
network: 
\begin{enumerate}
\item Unsynchronized or phase synchronized {\it non-homogeneous} 
behaviour : {\it phase one},

\item Partially synchronized or phase synchronized 
{\it homogeneous behaviour} : {\it phase two},
and 

\item Fully synchronized {\it homogeneous behaviour} : {\it phase three}.
\end{enumerate}
Here, synchronization refers to the variables at different nodes having the same
value
$x_i(t) = x_j(t)$ for all $i,j$. The network is globally synchronized when at 
each time $t$, 
all nodes have the same value. Partial synchronization means that some of the
nodes form a cluster inside which all the 
nodes are synchronized while they are not synchronized with the nodes in the
different clusters. We note, however, that this state usually does not 
occur in our
coupled dynamics because the phase differences between the various clusters 
will interfere with the internal synchronizations. The following behaviour, 
however, does robustly occur in suitable parameter regions. A pair of nodes 
is called phase synchronized \cite{PS1} when they have their minima (maxima) 
matching for all $ t > t_0$, that is, when one of them attains a minimum 
then so does the other. The concrete values may and can be different. In a
phase synchronized cluster all nodes are phase synchronized. 

Complete synchronization is indicated
by the  variance $\sigma^2$ of the variables over the network tending to zero, 
where
$$\sigma^2 = \left\langle \frac{1}{N-1} \sum_{i}[x_{i}
(t) - \bar{x}(t)]^{2}\right\rangle_{t},$$ 
 $\bar{x}(t)=\frac{1}{N}
\sum_{i}x_{i}(t)$ denotes an average over the nodes of the network, and $
\left\langle \dots \right\rangle _{t}$ denotes an average over time.
We define {\it homogeneous} and {\it non-homogeneous} behaviour based on 
the symbolic 
dynamics of the individual nodes. If all the nodes have the
same transition 
probabilities, then
we say that the coupled dynamics is homogeneous; otherwise it is 
non-homogeneous.
Homogeneity is indicated by the  variance of the transition
probability over the network being zero, 
\begin{equation}
\varsigma^2 = 
\left\langle \frac{1}{N-1}\sum_{k=1}^N[P_k(\alpha,\alpha)-
\overline{P(\alpha,\alpha)}]^2 \right\rangle\\
\label{equ-homo}
\end{equation}
where $\overline{P(\alpha,\alpha)}
= \sum_{k=1}^N P_k(\alpha, \alpha)$ denotes an average over the
nodes of the network.

\section{Homogeneous phases and coupled dynamics on network}
\subsection{Homogeneous phase and network properties} 
We shall now connect the {\it symbolic homogeneity} with 
network properties.
When all the nodes in a network have the same degree and the network is 
homogeneously connected, i.e. if the network is completely symmetric,
like a nearest neighbour coupled network with  periodic boundary conditions, 
then, unless the dynamics breaks the symmetry, each node should have 
{\it qualitatively the same symbolic 
dynamics}, i.e. all transition probabilities for all the nodes being
equal. In fact, one might then even expect stability of the
synchronized state, but that in general is not true for all coupling strengths.
Homogeneous symbolic dynamics need not correspond to synchronization,
though it may correspond to  phase synchronization. 
For random networks, homogeneity of the symbolic dynamics
depends on the number of connections in the network. 
 Note that depending upon the coupling strengths, for certain network
architecture one may get all the three
phases, including the homogeneous synchronized phase. Using the
master stability function which takes local dynamics as well
as network architecture into account, one can deduce the coupling strength
region for which the coupled dynamics would be synchronized \cite{MSF}.
To relate different
dynamical states with the network parameters, we consider the quantity
$r=\frac{N_c}{N(N-1)/2} \sim N_c/N^2$. This ratio compares 
the number of 
connections $N_c$ in the network
with the number of possible connections $N(N-1)/2$, 
We use $r$ as an indicator to roughly distinguish the three phases.
For $r$ being close to one (number of
connections $N_c$ of order $N^2$), we get a fully synchronized 
state for appropriate coupling strengths $\varepsilon$. Then 
the transition probabilities of all nodes are obviously equal (phase 3). For
$N_c \sim N $, we get phase two, i.e. the nodes are partially
synchronized or partially phase synchronized, but the symbolic dynamics 
of the nodes are identical. Note here we are only roughly
relating $N_c$ and phase, later we will provide a more accurate
relation between the number of connections and the phases.

\subsection{Symbolic synchronized phase and global properties of coupled %
dynamics}
\begin{figure*}
\includegraphics[bb=83 138 496 836,height=15cm]{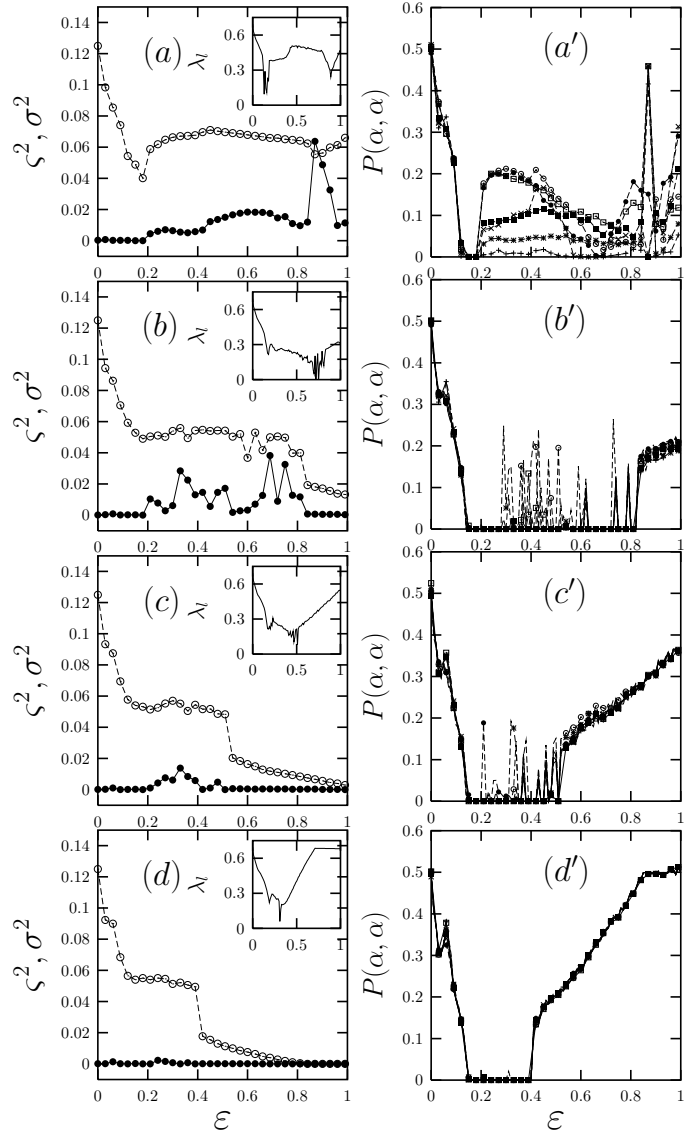}
\caption{Examples of coupled networks showing all three 
phases.
All figures are plotted for scale-free networks, generated by
using BA algorithm \cite{BA}, of size $N=200$ and 
(a) average degree 2 (phase 1),
(b) average degree 6, (c) average degree 10, (d) average degree 20. 
The $x$-axis represents the coupling strength and the $y$-axis gives 
the synchronization measure $\sigma^2$ ($\circ$) and the homogeneity measure
$\varsigma^2$ ($\bullet$) for the whole network. The largest Lyapunov
exponent is plotted as a function of the coupling strength (see inset).
Figs.~($a^\prime$), ($b^{\prime}$), ($c^{\prime}$) and ($d^{\prime}$) show 
exact values of the transition probability $P(\alpha, \alpha)$ for different 
nodes as a function of the coupling strength. For clarity we plot only a few
arbitrarily selected nodes.}
\label{sym-scale}
\end{figure*}

We concentrate on the phase where the nodes are not synchronized though 
their symbolic dynamics are identical. This is the most interesting phase 
as the complexity of the coupled dynamics can be understood
by observing the symbol sequence of any arbitrarily selected node.
Fig.~\ref{sym-scale} is plotted for the logistic map as the local map and a 
scale-free network 
\footnote{We have generated that scale-free network by the standard %
preferential attachment scheme \cite{BA}, but one should note that large %
classes of scale-free networks may exhibit a qualitatively different behaviour %
as regards other crucial network parameters besides the degree sequence, in %
particular concerning synchronizability, see \cite{ABJ}.} as the coupling 
network. Figs.~(a),(b),(c) and (d) plot the variation of synchronization 
measure ($\sigma^2$), and measure of homogeneity ($\varsigma^2$) as a function
of coupling strengths and Fig.~($a^\prime$), ($b^\prime$),
($c^\prime$) and ($d^\prime$) plot the  
transition probability $P(\alpha, \alpha)$ for different nodes.   
We start with the example of networks having coupled dynamics in phase one 
(non-homogeneous unsynchronized) and we move towards the examples
of networks showing
homogeneous synchronized state, phase 3. 

For $\varepsilon < 0.2$, the coupled logistic map model (\ref{coup-dyn}) 
exhibits a similar behaviour for different  coupling architectures, 
with quasiperiodic behaviour around $\varepsilon=0.2$. The 
behaviour varies with the coupling architecture for larger coupling 
strengths. In the periodic regions the symbolic dynamics of the nodes, 
given by (\ref{sym}), would always be similar irrespective of the underlying 
coupling network. So, in the periodic regions we do not get any extra
information about the network by observing symbolic sequences, but
the symbolic dynamics is informative when the coupled dynamics lies on the
chaotic attractor. Subfigures~\ref{sym-scale}(a), $(a\prime)$, are plotted 
for scale free networks with average degree 2. The transition probabilities 
$P(\alpha, \alpha)$ are completely different for the different nodes 
(except for $\varepsilon < 0.2$). Here, the nodes are not
synchronized, $\sigma^2$ being nonzero for the entire coupling strength
range. This state corresponds to {\it phase one}. Interesting phenomena
occur when we increase the number of connections in the networks. 
Subfigures~\ref{sym-scale}(b) and ($b^\prime$) are plotted for a 
scale-free network
with  average degree 6. It can be seen that  $P(\alpha,\alpha)$ for 
different nodes are 
remarkably similar. Note that we calculate $P(\alpha, \alpha)$ for 
coupled dynamics being in the chaotic and unsynchronized regime
($\lambda_l$ and $\sigma^2$ both are greater then zero). So we do not
take periodic and synchronized regions into account 
which obviously yield similar symbolic dynamics for all the nodes.
This homogeneity becomes more prominent as we increase the number of 
connections in the network. In Figs.~\ref{sym-scale}(c), ($c^\prime$) and 
\ref{sym-scale} (d), ($d^\prime$),
The transition probabilities of all the nodes are the same except for a 
few places where some nodes have different transition probability 
(e.g.~node number 50 in ($c^\prime$) having a different value of 
$P(\alpha, \alpha)$). Note
that here the nodes are not synchronized, which is indicated by the
nonzero value of $\sigma^2$ throughout the coupling range, except 
for $\varepsilon=1$ in (c) and for $\varepsilon > 0.8$ in (d).

The second interesting feature is  
the qualitatively similar behaviour of the
largest Lyapunov exponent, which is calculated from ~(\ref{coup-dyn}),
and $P(\alpha, \alpha)$, which is calculated from a scalar time series
of an arbitrarily selected node. Note that the  time series used for
the calculation of $P(\alpha, \alpha)$ is very short compared to the 
traditional methods to calculate the largest Lyapunov exponent from a scalar
time series. This similar behaviour of the Lyapunov exponent
and the ordering relations between the values of the state variable  was 
first observed by Bandt and Pompe \cite{PE1} in the case of isolated
dynamics. Here we show that a similar relation exists for 
coupled dynamics, depending upon the network 
parameters, namely the connection architecture and the connection ratio 
$N_c/N^2$. 

\begin{figure*}
\includegraphics[bb=60 455 402 780, width=10cm,height=8cm]{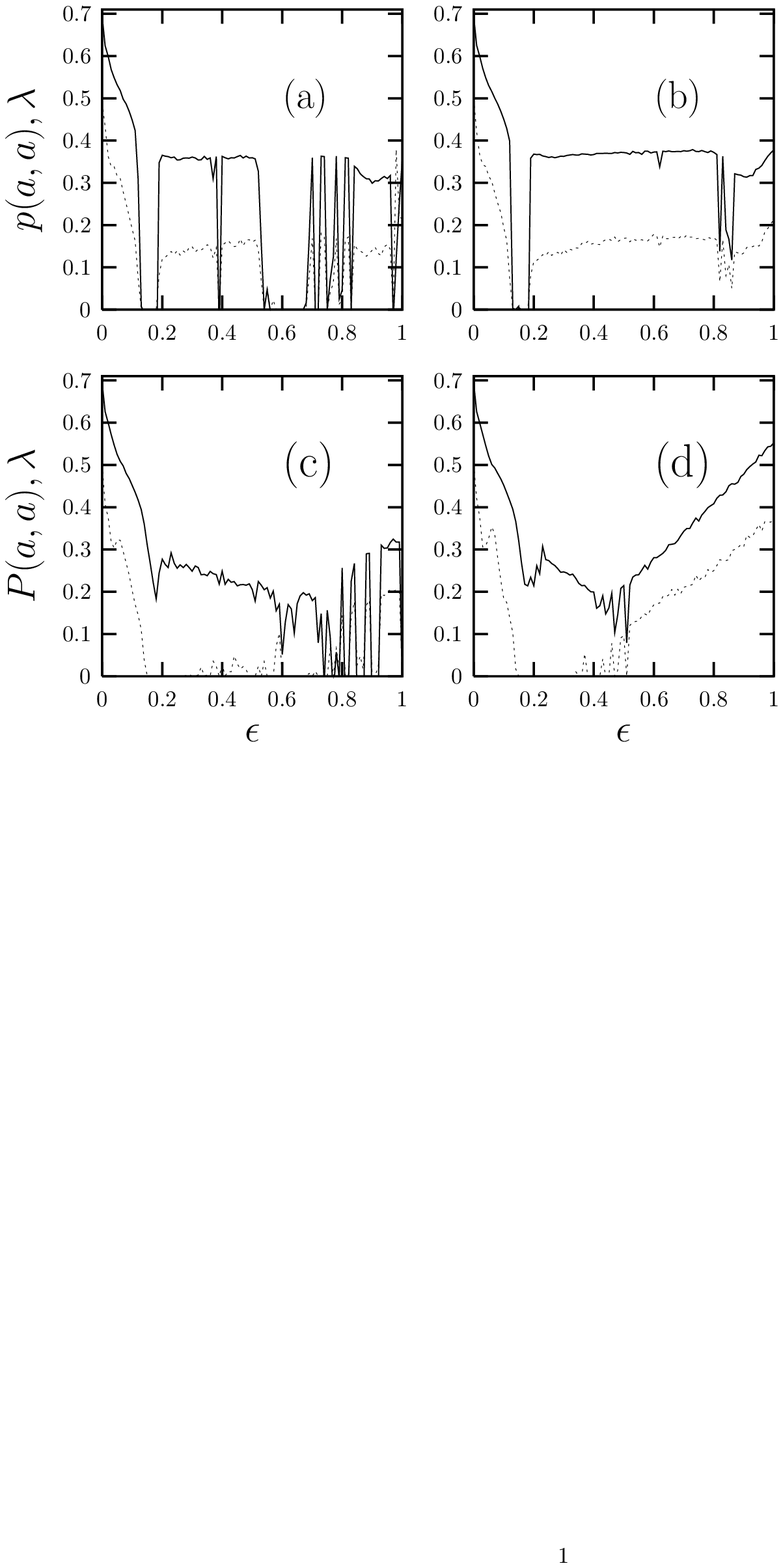}
\caption{The global measure of coupled dynamics from the local symbolic 
dynamics. We take various networks having coupled dynamics in the 
{\it phase two}. The $x$-axis gives the coupling strength $\varepsilon$ and 
the $y$-axis depicts $\lambda_l$ (-)(largest Lyapunov exponent for the 
coupled dynamics) as well as $P(\alpha, \alpha)$ (...)(transition probability 
for a randomly selected node). (a) for nearest neighbour coupled network of 
size $N=20$, (b) for 3-nearest neighbour coupled network, $N=50$, (c) and 
(d) are for random and scale-free networks, respectively, with average 
degree 10 and $N=200$.}
\label{sym-lya}
\end{figure*}

Fig.~\ref{sym-lya} is plotted for various networks being in the
{\it phase two} (homogeneous unsynchronized phase). They show
the similar behaviour of $\lambda_l$ and 
$P(\alpha, \alpha)$ of any arbitrarily selected node. 
For $k$-nearest neighbour coupled networks we always find the  
homogeneous phase,
independent of the average degree or the ratio $N_c/N^2$. This is 
because of the symmetry between the nodes.
In all the subfigures, $P(\alpha, \alpha)$  qualitatively matches with the 
largest Lyapunov exponent of the coupled dynamics. 
At certain $\varepsilon$ values $P(\alpha,\alpha)$ is
almost zero, whereas $\lambda_l$ is positive. This indicates ordered
behaviour (Ref. \cite{PS1}) of the  coupled system, for example Fig.2(c) 
and Fig.4(c) are plotted for the same network. When 
$P(\alpha,\alpha)$ is very small the coupled dynamics is in the high state of 
the phase synchronization (Fig.4(c)) (i.e. almost all nodes are
forming phase synchronized cluster(s) Ref.\cite{PS1}) though the largest
Lyapunov exponent still remains positive (Fig.2(c)).

In {\it phase three}, which is synchronized phase,
the largest Lyapunov exponent would simply be the 
Lyapunov exponent of the uncoupled map and all the nodes will have the same
transition probability as of the uncoupled node. 

\section{Relation between dynamical phases and network properties}
\begin{figure}
\includegraphics[bb=67 619 252 780]{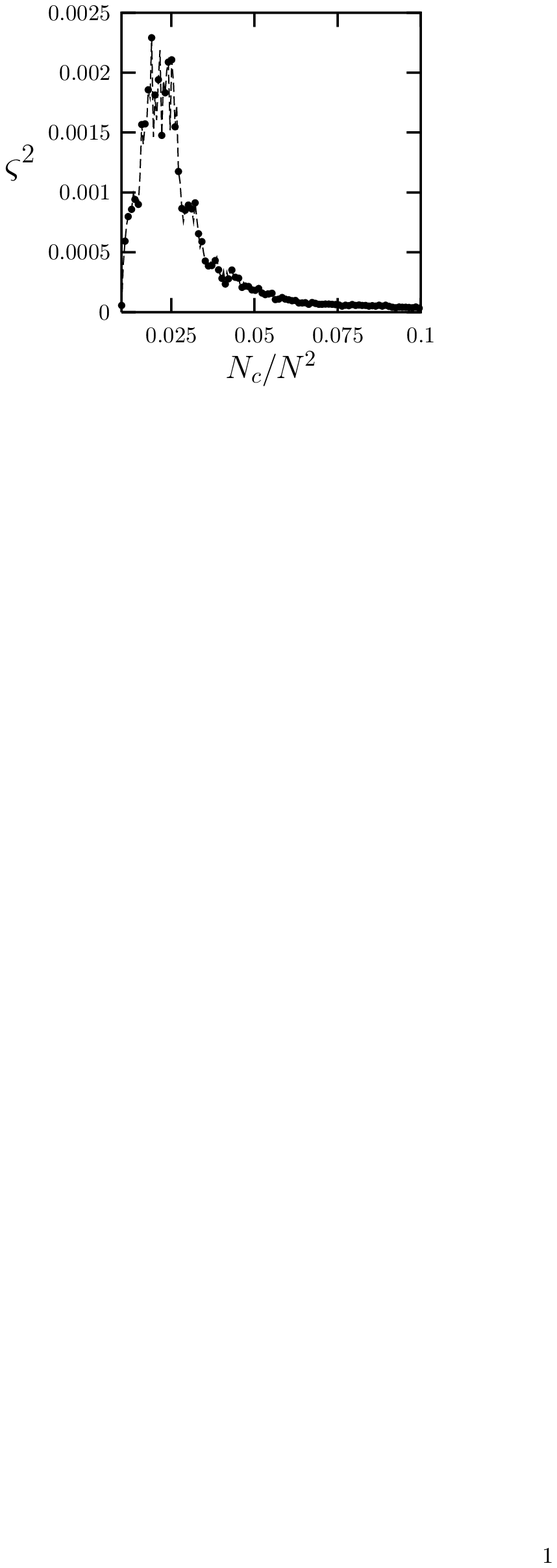}
\caption{The measure of 
homogeneity $\varsigma^2$ as a function of the connectivity ratio $2 N_c/N(N-1)$.}
\label{sym-nc}
\end{figure}

We can also exhibit a direct relation between network parameters and 
dynamical behaviour. The symmetry properties  of the network 
and the connection density affect the homogeneity
of the symbolic sequences.
Fig.~\ref{sym-nc} plots the deviation from the homogeneity 
indicated by $\varsigma^2_{\varepsilon} = 
\left\langle \frac{1}{N}\sum_{i=1}^N[P_i(\alpha,\alpha)-
\overline{P(\alpha,\alpha)}]^2 \right\rangle_{\varepsilon}$,
as a function of $2N_c/N(N-1)$. Here, $P_i$ is the transition probability of $i$th
node and $\overline{P(\alpha,\alpha)} = 
\frac{1}{N} \sum_{i=1}^N P_i(\alpha, \alpha)$, and
$\left \langle \cdot \right \rangle_{\varepsilon}$ denotes the average over all coupling 
strengths.
We start with one-dimensional nearest neighbour coupled
network (homogeneous phase) and randomly add connections. For
nearest neighbour coupled networks we obtain the homogeneous phase, as already 
explained in the previous section. As we add the connections randomly, 
first the homogeneity gets perturbed, but  gets 
reestablished as the number of connections is increased further. Note that
here we always calculate the deviation in the non-synchronized regime only, 
because the synchronized regime obviously corresponds to the homogeneous phase.
For each randomly added connection we take the average of the twenty networks.
Note that in this region (phase two) $\sigma^2$ is not zero.
For $N_c/N^2$ close to one, we get a synchronized
state after a coupling strength \cite{SJ-analytic} that corresponds to the
homogeneous state (phase three). 
Note that for the local dynamics in the chaotic regime, only the network 
property seem to be responsible for the homogeneous or non-homogeneous 
behaviour of the coupled dynamics. Fig.~\ref{sym-nc} is plotted for the 
coupled logistic map but a similar behaviour is shown by the H\'enon maps 
also (see section VII). 

\section{Phase synchronization : Symbolic synchronization}
\begin{figure*}
\includegraphics[bb=61 369 502 780,width=14cm,height=10cm]{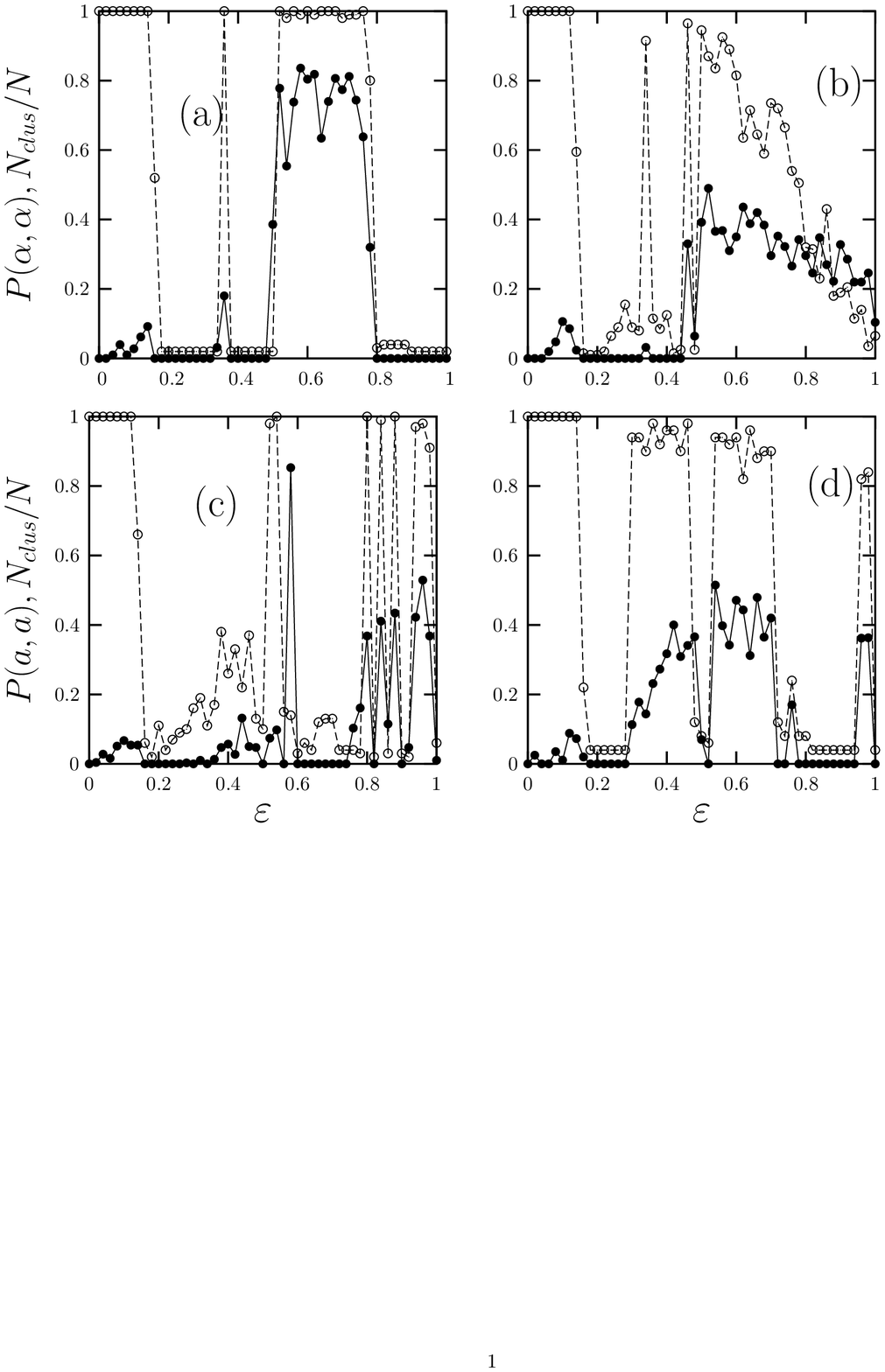}
\caption{The ratio of the number phase synchronized clusters to
the maximum possible clusters, $N_\text{clus}/N$ and
the transition probability $P(\beta, \beta)$ for the coupled logistic map
as a function of the coupling strength
$\varepsilon$, (a) for a nearest neighbour coupled network with  average 
degree 20 and $N=100$, (b) for  a scale-free network with average degree 10 and
$N=100$, (c) for a random network with average degree 10 and $N=100$,
(d) for  a nearest neighbour coupled network with average degree 6 and $N=50$, 
and the tent map $(f(x)=(1-2|x-\frac{1}{2}|)$ as the local chaotic function.}
\label{sym-phase}
\end{figure*}

If nodes $i$ and $j$ have the same symbolic dynamics, $s_t(i) = s_t(j)$, then 
we say nodes $i, j$ are symbolically synchronized. Also, a cluster of nodes 
is symbolically synchronized if all pairs of nodes belonging to that cluster 
are symbolically synchronized. Note that in a symbolically synchronised 
cluster, the state values of the nodes may differ. The symbolic 
synchronization is observed in the {\it phase two}, where the number
of the connections in the networks is very small, in general of the order
of $N$. With the increase in the number of connections we usually get a 
fully synchronized cluster, which  is trivially symbolically synchronized.
Many real-world networks are sparsely connected $(N_c \sim N)$,
and complete synchronization is relatively rare, though phase synchronization or 
symbolic synchronization is possible. We show that $P(\beta, \beta)$ can 
be used as a good measure of the phase synchronization in the coupled map 
network (\ref{coup-dyn}). Fig.~\ref{sym-phase} shows the correlation between 
the  phase synchronization and the transition probability $P(\beta,\beta)$ 
of an arbitrary selected node. 
We see that in the homogeneous region $P(\beta, \beta)$ matches 
considerably well with the phase synchronization. Note that in partially 
ordered phase region, phase synchronized clusters vary with time. We plot 
the number of clusters calculated for a certain time length, and the number 
of clusters may change with the evolution of the coupled dynamics. Therefore 
at some coupling strength region(s), the transition probability 
$P(\beta, \beta)$ does not match with the phase synchronization 
(For example in the Fig.~\ref{sym-phase}(c), at coupling strength 0.59,
the value of $P(\beta, \beta)$ is very high although the nodes are phase 
synchronized).

\section{Coupled H\'enon maps}
In this section we apply our method to coupled H\'enon maps. 
The H\'enon map is a two-dimensional map
\cite{henon},
\begin{eqnarray}
x(t+1) = y(t) + 1 - a x(t)^2 \nonumber \\
y(t+1) = b y(t). \nonumber
\end{eqnarray}
When one introduces the possibility of a time delay, the above equation 
can be written as the scalar equation,
\begin{equation}
x(t+1) = b x(t-1) + 1 - a x(t)^2
\label{henon-map}
\end{equation}
For the parameters we take the values
$a = 1.4$ and $b = 0.3$, for which the H\'enon map
is known to have a chaotic attractor. 

\begin{figure}
\centering
\includegraphics[width=\columnwidth]{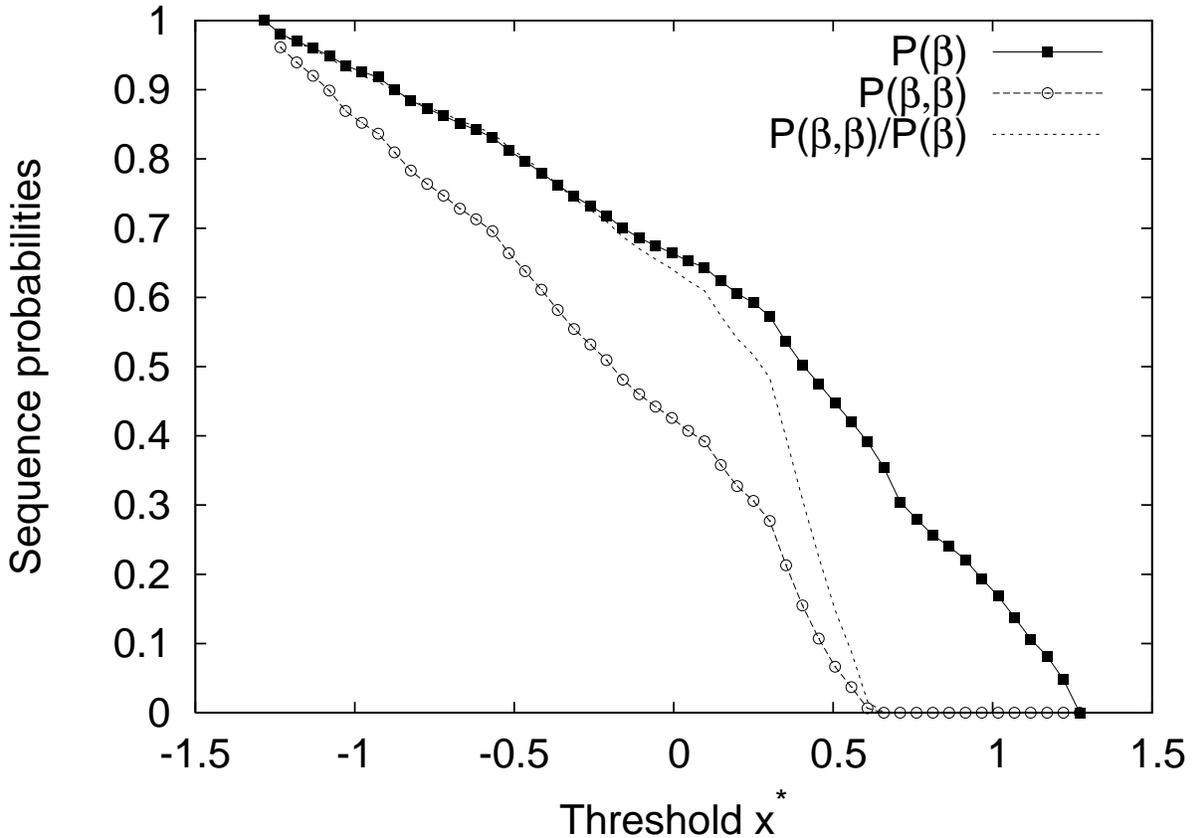}
\caption{Illustration of the choice of the threshold $x^{\ast}$
as the point where $P(\beta,\beta)/P(\beta)$ sharply drops to near zero. }
\label{fig:henonx}
\end{figure}

We define the symbolic dynamics as given in (\ref{sym}).
The choice of the threshold $x^{\ast}$ requires some care.  
A judicious choice should make certain short transition
probabilities very small, which may be useful for detecting network dynamics
from single-node measurements \cite{pap1}. Clearly,
increasing the threshold decreases the probability of occurrence of the
repeated sequence $\beta\beta$. However, it also decreases the probability of 
observing
the single symbol $\beta$, making it difficult to work with short time series.
Hence, the choice of the threshold is a compromise between these two effects.
We use the natural density defined by the data to choose a threshold.
Fig.~\ref{fig:henonx} depicts how the the probabilities of observing
a single symbol $\beta$ and the repeated sequence $\beta\beta$ 
change depending on the value
of the threshold $x^{\ast}$. It can be seen that a choice of $x^{\ast}$
roughly in the range $(0.55,1.20)$ would be useful, since it renders the
sequence $\beta\beta$ very unlikely without constraining the occurrence of 
the symbol $\beta$. Note that it is immediate from their definitions that 
the probabilities $P(\beta)$ and $P(\beta,\beta)$ will be decreasing as 
functions of $x^{\ast}$, and will approach zero as $x^{\ast}$ increases; 
furthermore, $P(\beta)>P(\beta,\beta)$. It follows that one can find a 
threshold $x^{\ast}$ for which $P(\beta)$ is large compared to 
$P(\beta,\beta)$. Fig.~\ref{fig:henonx} shows the ratio 
$P(\beta,\beta)/P(\beta)$, and the sharp decrease at about 
$x^{\ast}\approx 0.6$ suggests to take some value near 0.6 as the threshold, yielding a
very small $P(\beta,\beta)$ and a large $P(\beta)$ at the same time.
We evolve (\ref{coup-dyn}) starting from random initial conditions, with
(\ref{henon-map}) as local dynamics,
and estimate the transition probabilities $P(i, j)$ as discussed in
the first section, using a time series of length
$\tau =1000$. 
At the globally synchronized state 
$x_{i}(t)=x_{j}(t); \,  \forall i,j, t$, with all
nodes evolving according to the rule (\ref{henon-map}), the symbolic sequences
measured from a node will be
subject to the same constraints as that generated by (\ref{henon-map}).

\begin{figure*}
\includegraphics[bb=57 418 585 780,width=15cm]{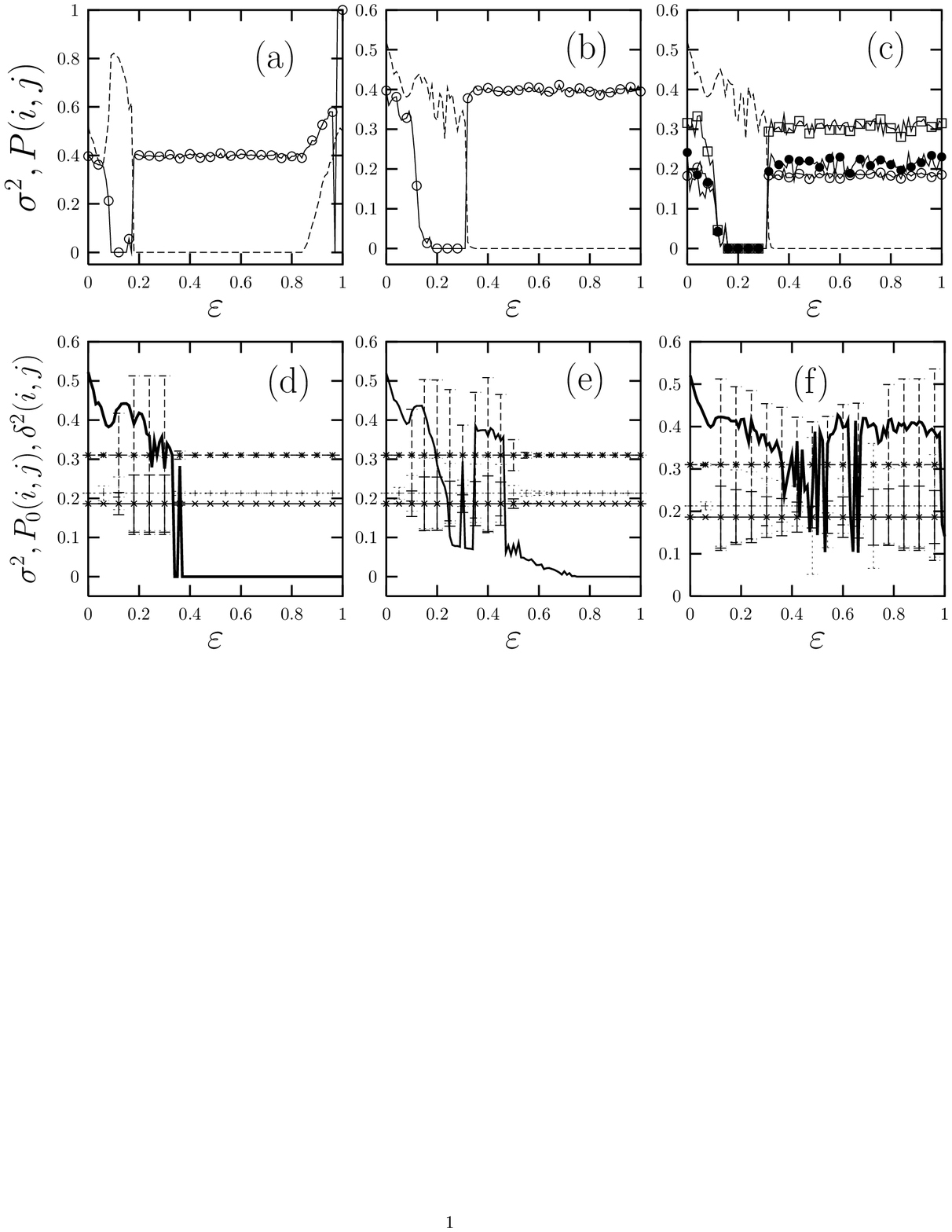}
\caption{The transition probability measure for 
coupled H\'enon maps. The $x$-axis displays the coupling strength and the 
$y$-axis shows the different
transition probabilities and the measure of synchronization $\sigma^2$.
(a) is for two coupled nodes and  
plots $P(\alpha, \alpha)$ for the
symbolic sequence of length 2. (b) and (c) are plotted for globally coupled 
networks with $N=50$. (b) plots $P(\alpha, \alpha)$  
and (c) plots transition probabilities for  symbolic sequences of  length 3,
namely $P(\alpha, \alpha, \alpha)$ ($\square$), $P(\alpha, \alpha, \beta)$ 
($\bullet$), and $P(\beta, \alpha, \alpha)$ ($\circ$). 
The synchronized
state is detected when all the transition probabilities are equal to those
of the uncoupled map; i.e. the transition probabilities at
the zero coupling strength. (d), (e) and (f) show the standard deviation
$\sigma^2$ (solid thick line) and  
$\delta^2$ (vertical dashed line) for 
these three transition probabilities of an arbitrary selected node with
respect to the transition probabilities of the uncoupled function 
(solid line), i.e for $\varepsilon=0$. 
$\delta^2$ is calculated for 20 simulations for the dynamics with different
sets of random initial conditions. (d) for 
a globally connected network with $N=50$, (e) and (f) for a random network 
with $N=100$, average degree 10 and 2 respectively. The last subfigure is 
plotted to show the behaviour of the transition probabilities when we do 
not get global synchrony even at large coupling strengths. }
\label{fig1-henon}
\end{figure*}

We now discuss some results based on  numerical simulations
on various networks. Fig.~\ref{fig1-henon} plots the 
transition probabilities as a function of the coupling strength.
We consider the symbolic sequences of length two and  three.
For length two, we consider the  transition probabilities  
$P(\alpha, \alpha)$ and $P(\beta,\beta)$. For sequences of length three we 
have 6 possible transitions, but some of them are very small 
(like $P(\beta,\beta,i)$, $i$ being $\alpha, \beta$ ), so we plot only 
those transition probabilities which vary with the couplings.
It is clear from the figures that the synchronized state is easily 
detected by looking at the transition probabilities of any arbitrarily 
selected node. Whenever the transition probabilities are equal to the
transition probabilities of the map (\ref{henon-map}), the network is 
globally synchronized.
It is clear from subfigures (d),(e) and (f) that for the synchronized region
(zero $\sigma^2$), the deviation of transition probabilities from the 
transition probabilities of the uncoupled map (\ref{henon-map}) is
also zero.
Here, the deviation of $P(i,j)$ of any node is defined as 
$\delta^2_{i,j} = \left\langle \frac{1}{m-1}\sum_{k=1}^{m}[P_k(i,j)-
\overline{P(i, j)}]^2]\right\rangle$, where 
$\overline{P(i, j)} = \frac{1}{m}\sum_k P_k(i,j)$, calculated at $\varepsilon=0$
$k = 1, \dots \dots $ are $m$ different sets of random initial conditions
taken between $-1.5$ and $1.5$.
In all the figures (except (f)) we get synchronization for  larger
coupling strengths, so the deviation is almost zero there, i.e. all 
transition probabilities match
completely with those of the uncoupled map. Note that there are certain regions
(small coupling strength range $\varepsilon < 0.2$) where the nodes do not 
get synchronized while the deviations are quite small. That is because for 
sufficiently small coupling strength,  couplings 
do not affect the behaviour of the individual nodes very much, and so the
transition probabilities do not differ much from those 
corresponding to the uncoupled function.
However, as we increase the coupling strength, the transition probabilities
 become dependent on the couplings. 
Still, if we look at the small coupling strength regions carefully
we see that not all the deviations are small. For example, although the 
deviations of $P(\beta, \alpha, \alpha)$ (- - -)
and $P(\alpha, \alpha, \beta)$ (-) are very small, the deviation in
$P(\alpha, \alpha, \alpha)$ ($\dots$) is still large, whereas for the
synchronized regime all deviations are very close to zero.
\begin{figure}
\includegraphics[bb=67 619 252 780]{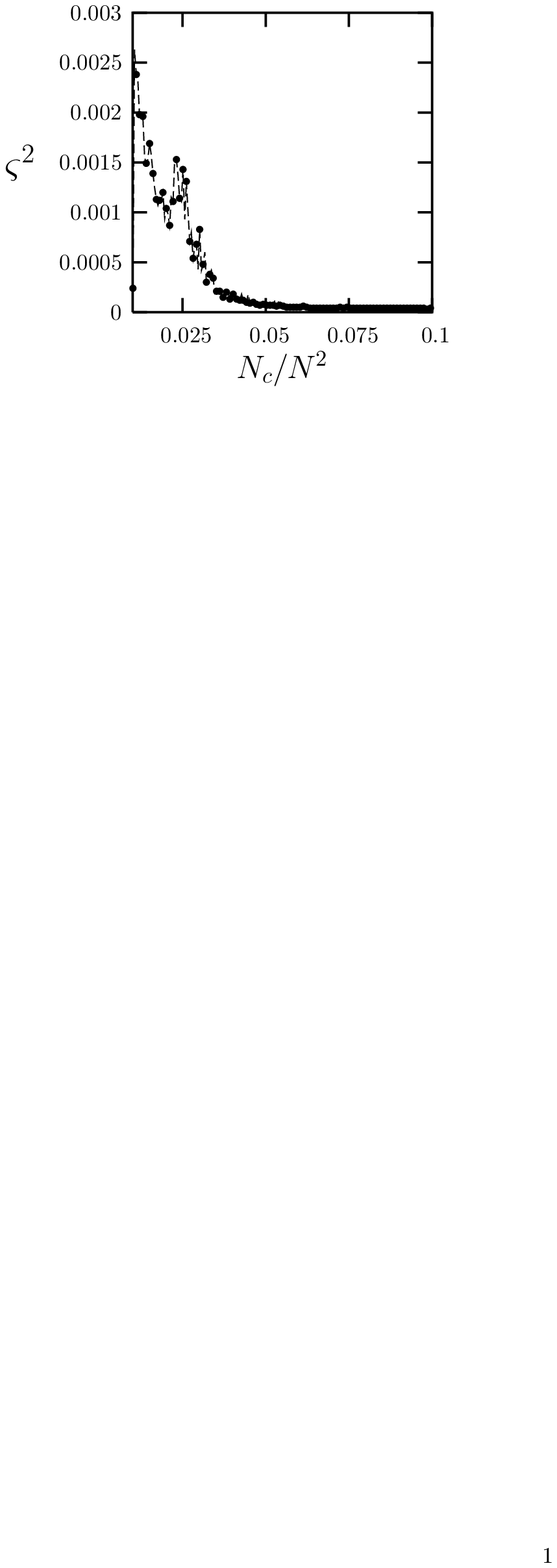}
\caption{The measure of
homogeneity $\varsigma^2$ as a function of connection ratio $2 N_c/N(N-1)$,
with H\'enon map as the local dynamical function.}
\label{sym-nc-henon}
\end{figure}
Fig.~\ref{sym-nc-henon} plots the deviation from the homogeneity
$\varsigma^2$ as a function of $N_c/N^2$ (see the caption
of Fig.~\ref{sym-nc}, which shows 
a similar plot with logistic map as a local dynamical function).
The only difference is that 
for the H\'enon map we plot the transition probability of three-symbol 
sequences instead of two-symbol sequences for logistic and tent maps.

\section{conclusion}
We have studied the symbolic dynamics of coupled maps on networks. We define 
our symbolic dynamics based on non-generating partitions leading to some 
forbidden transitions of symbols in the time evolution of the function. The
optimal partitions are those which lead to the maximal difference
between the permutation entropy of the dynamical iteration and corresponding
random iteration. For one-dimensional systems finding these partitions
is simple, whereas for higher dimensional systems it may be more difficult. 
However, it turns out that symbolic dynamics drawn from any non-generating 
partitions is usually good enough for the applications we have considered in 
this paper. The symbolic dynamics can be drawn when the system parameters are 
not known, as well as for experimental data taken in a noisy environment.

We use  symbolic dynamics as a measure of dynamical state of the coupled 
system and show various applications of this measure.  
We define three different states of the coupled dynamics based on 
the synchronization and the symbolic dynamical properties. 
In the homogeneous synchronized phase, complete synchrony is detected by 
comparing the transition probabilities of any arbitrarily selected node 
with those of the uncoupled function. In this state the coupled dynamics 
collapses to the dynamics of the uncoupled function, and the symbolic 
dynamics of any node is subject to the same constraints as that generated by 
the uncoupled function.

Phase two, which 
refers to the {\it homogeneous unsynchronized phase} or 
{\it symbolic synchronized phase}, is of our prime interest where the nodes 
are not synchronized but have identical  symbolic dynamics.
Although these phases are detected dynamically, we find that the homogeneous
unsynchronized phase is related to the connection density ($N_c/N^2$)
 and to a smaller extent
to the chaotic dynamical function used. This region is generally observed for 
networks with $N_c \sim r \times N^2$ where $0.05 <  r < 0.1$. Most of the 
real networks are sparsely connected and come under the category of phase 
two. In this phase we can deduce the global properties of the coupled 
dynamics such as the largest Lyapunov exponent and phase synchronization by 
simply observing the local symbolic dynamics of any randomly selected node. 

As it is expected from such a 
simplistic reduction, our symbolic dynamics gives only the qualitative 
understanding. A more precise calculation of complexity through some entropy 
measure of the system based on the symbolic dynamics is one of the future 
steps. Further future investigations will involve an analytical understanding 
of symbolic synchronization and application to detect various levels of 
synchronization in experimental data taken from  
coupled systems.

\end{document}